\documentclass[aps, twocolumn, 10pt, pra, superscriptaddress]{revtex4-1}
\usepackage{amsmath} 
\usepackage[utf8]{inputenc}
\usepackage[T1]{fontenc}
\usepackage{color}
\usepackage{xcolor}
\usepackage{graphicx}
\usepackage{titleref}

\usepackage{extarrows}
\usepackage{amssymb}
\usepackage{upgreek}

\begin{document}

\preprint{aps/123-qed}
%\title{In-situ detection of crystallisation from liquid solutions and precise temperature control}
\title{In-situ detection of nucleation in high temperature solutions}
\author{Andreas G. F. Schneider}
\affiliation{EP VI, Center for Electronic Correlations and Magnetism, Institute of Physics, University of Augsburg, D-86159 Augsburg, Germany}
\author{Paul Sass}
\affiliation{ScIDre - Scientific Instruments Dresden GmbH, Gutzkowstr. 30, D-01069 Dresden,
Germany}
\author{Robert Schöndube}
\affiliation{ScIDre - Scientific Instruments Dresden GmbH, Gutzkowstr. 30, D-01069 Dresden,
Germany}
\author{Anton Jesche}
\email{anton.jesche@physik.uni-augsburg.de}
\affiliation{EP VI, Center for Electronic Correlations and Magnetism, Institute of Physics, University of Augsburg, D-86159 Augsburg, Germany}

\date{\today}

\begin{abstract}
The state of a sample during crystal growth from high temperature solutions is not accessible in conventional furnace systems.
An optimization of the growth parameters often requires arduous trial and error procedures in particular in case of novel multicomponent systems with unknown phase diagrams.
Here we present a measurement technique based on lock-in amplification that allows for in-situ detection of the liquidus and solidus temperatures as well as structural phase transitions.
A thin, metallic measurement wire is mounted in close vicinity to the melt. 
Characteristic anomalies in the time-dependent electrical resistivity of this wire allow for the detection of latent heat release without using a reference crucible. 
The method is implemented in a 'feedback furnace' and enables an adjustment of the temperature profile based on the occurrence or absence of phase transitions.  
The absolute temperature serves as an additional source of information. 
%In contrast to conventional procedures, the state of a sample is identified not only by the nominal temperature in the process chamber, but by the onset or absence of characteristic signatures, which are detected in real time.
Obtained phase transition temperatures are in good agreement with differential thermal analysis (DTA).  
%directly originating in the crystallization processes of the sample material.\\
%Measurements on exemplary alloys are given in to comparison with conventional DTA measurements.
% The latter show a distinct comparability to this very common and established technique.
\end{abstract}

\maketitle
\section{Introduction}
High-temperature solutions, also referred to as flux, provide a powerful tool for the single crystals growth of binary or higher multicomponent compounds\,\cite{Elwell1975,Canfield2001, Flusszuechtung, Jesche2014c}. 
A melt composed of several elements is deliberately cooled such that crystallization of certain fractions of the melt and growth of monocrystalline grains of the desired phase occur. 
Problematic and inaccurate is the prediction of the temperature at which the sought-after phase nucleates, as the phase transition temperatures are rarely known in the case of novel multicomponent systems. 
Thus, after melting and homogenization of the sample material, the mixture is cooled as slowly as achievable over the critical temperature (liquidus temperature) determined from the phase diagram (if available) or from previous investigations. 
This procedure has the consequence that the exact beginning of the crystallization cannot be determined and the slow cooling of a melt solution usually has to be started 50$^{\circ}$C to 100$^{\circ}$C above the nominal liquidus temperature. 
With cooling rates of 2$\frac{^{\circ}\text{C}}{\text{h}}$, the process can last several days before the actual onset of nucleation and crystal growth take place. 
This means an enormous expenditure of time and energy, which can be even further increased by supercooling of the melt. 
In the temperature range close to nucleation and the early phase of crystal growth, however, even smaller cooling rates of less than 2$\frac{^{\circ}\text{C}}{\text{h}}$ would be desirable in order to allow undisturbed formation of separated crystals without intergrowth. 

%In conventional furnaces, nucleation stays undetected and a reaction to the onset of crystallization is not possible since the exact point in time remains undetected.
Here, we present a novel method for controlling the heating power of a furnace. 
The key idea is that the state of a sample is no longer determined only by the nominal temperature in the process chamber, but by the onset or absence of characteristic signatures in the time-dependent temperature measured directly at the sample position.
Those signatures are caused by the latent heat associated with first order phase transitions.
Latent heat anomalies in the time-dependent sample temperature are detectable by high-precision measurements of the electrical resistivity using a lock-in-technique. 
We succeeded in detecting temperature anomalies smaller than $10^{-4}$ of the absolute value by phase-sensitive measurements in dilute alloys. 
The in-situ detection of nucleation makes it possible to cool below or oscillate around the liquidus temperature at significantly slower rates than previously practical.

The paper is organized as follows: we specify our furnace and measurement setup followed by further experimental details on materials and equipment. 
In the second part, we present results on binary alloys: We show precise liquidus and solidus temperature detection of Bi-rich and In-rich solutions, investigate the development of the characteristic signatures upon successive dilution for Ni-Bi and describe detailed measurements of several phase transitions in Pd-Bi solutions.

%Furnace setup and resistivity measurements are presented in Sec.\,\ref{setup} followed by further experimental details in Sec.\,\ref{exp}. 
%The detection of liquidus and solidus temperature is shown in Sec.\,\ref{bina} for Bi-rich (Ni-Bi, Mn-Bi, and Pd-Bi) and In-rich solutions (Sb-In, Au-In, and Pd-In).
%The development of the characteristic signatures upon successive dilution were investigated for Ni-Bi (Sec.\,\ref{dilute}).
%Detailed measurements of several phase transitions including a structural one were performed for PdBi (Sec.\,\ref{pdbi}).
%A Discussion follows in Sec.\,\ref{disc}.

\section{Furnace setup}\label{setup}

\begin{figure}
\begin{center}
\includegraphics[width=0.9\columnwidth]{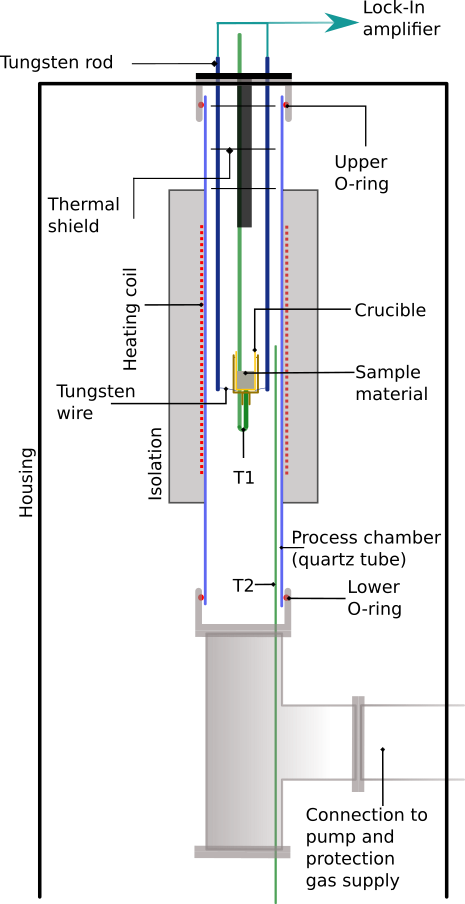}
\caption{Schematic sketch of the feedback furnace. 
Measurement wire (tungsten) and thermocouple (T1) are directly attached to the growth crucible, which is placed in an inert-atmosphere chamber.
}
\label{Gesamtaufbau}
\end{center}
\end{figure} 

We implemented the feedback method in in custom-build furnace produced by ScIDre\,\cite{scidrewebsite}. 
The tubular furnace consists of a quartz process chamber (wall thickness 2.5\,mm, diameter 50\,mm, length 350-360\,mm), which is centrally wrapped with a FeCrAl heating wire (diameter 1\,mm, length 180\,mm).
%The furnace is a tubular furnace (Fig.\,\ref{Gesamtaufbau}) whose reaction chamber consists of quartz tube (wall thickness 2.5\,mm, diameter 50\,mm, length 350-360\,mm), which is centrally wrapped with a heating wire (FeCrAl, diameter 1\,mm, length 180\,mm).
Each end of the quarz tube is plunged in a flange made of stainless steel and sealed by an O-ring (FKM 80A). 
The lower flange is connected to a vacuum pump and a protection gas supply while the upper flange gets closed by a cover plate with several gas-tight feedthroughs. 
The setup features two thermocouples, one lead into the furnace by the upper side (Fig.\,\ref{Gesamtaufbau} $\rightarrow$ T1) and one by the bottom side (Fig.\,\ref{Gesamtaufbau} $\rightarrow$ T2). 
T1 is fed through the cover plate reaching into the center of the heating zone, where the crucible is located. 
To this extend, the thermocouple is bent upwards by 180$^\circ$ and the crucible is placed on top. 
The crucible is placed slightly below the center of the heating zone for providing a temperature gradient of 20$\frac{^{\circ}\text{C}}{\text{cm}}$, measured by successively shifting the position of the thermocouple.
Note that this value presents an upper limit since thermal conductivity and convection in the melt reduce the temperature gradient.
%By the thin isolation the gradient reaches 20$\frac{^{\circ}\text{C}}{\text{cm}}$ along the crucible.\\
Whereas T1 is used to measure the absolute temperature as close as possible to the sample material, T2 serves primarily as safety measure for avoiding uncontrolled heating in case of faulty T1 readings.
All processes are performed under Argon atmosphere at a pressure of roughly 500\,mbar.

%NEUER ABSATZ
The crucible preparation deserves special attention as it is a major component of our phase transition detection method. 
In our setup, the actual resistivity measurement and thus the detection of latent heat anomalies takes place within the bottom of the crucible. 
In order to detect temperature changes of the sample material as precise as possible, a thin measurement wire has to be placed as close as possible to the melt without direct contact, covering as big part of the melt surface as possible. 
Therefor, we use a tungsten wire of typically $40\,\upmu$m diameter which is placed in a meandering way inside the crucible bottom (Fig. 2). 
Several grooves are sawn into the bottom of the crucible and the measurement wire is covered by a ceramic glue.
A shallow indentation is drilled afterwards into the bottom of the crucible were there thermocouple T1 is placed upon crucible installation within the process chamber.

The filled crucible is loosely covered with an alumina cap and placed in the furnace under Argon atmosphere at a pressure of \mbox{500 mbar}. 
A piece of Zr-foil was placed on top as oxygen getter.
Note that materials with elevated vapor pressure could transport and deposit outside the crucible. 
Growth attempts with Bi-rich flux indicate that vapor pressures of $\approx 3$\,mbar at $T = 1000^{\circ}$C\,\cite{Fischer1966} do not lead to measurable mass loss (neglecting transport active species).

%Grooves are sawn into the bottom of the crucible at a distance of roughly 1\,mm by using a diamond multiwire saw (Fig.\,\ref{Tiegelbearbeitung}). 
%Furthermore an indentation \mbox{(diameter $\approx 3.5$\,mm)} is drilled into the bottom side where the crucible is placed on T1. 
%The measurement wire (tungsten, diameter 13-42\,$\mu$m) is placed in the grooves in a meandering way and covered by a ceramic glue. 
%Wire and thermocouple are placed as close as possible to the melt without direct contact.
%The readings of T1 are used to convert the voltage drop at the measurement wire to absolute temperature. 
%Also metallic crucibles can be used (not shown here), which require a different contacting method, that electrically insulates the wire and the crucible from each other. 
%At each side of the crucible stays a surplus of \mbox{50-70 mm}. 

\begin{figure}
\begin{center}
\includegraphics[width=0.8\columnwidth]{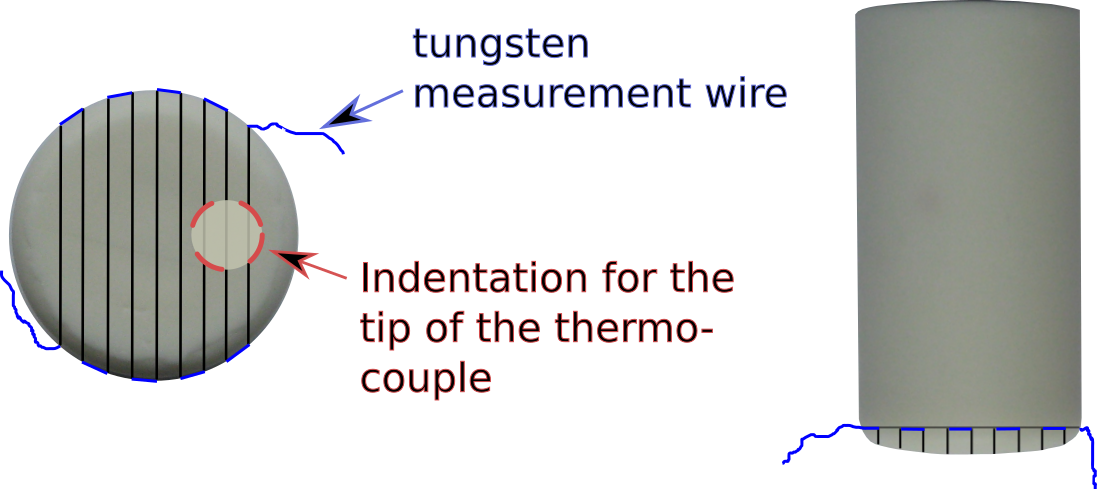}
\caption{Growth crucible. The measurement wire is mounted in a meandering fashion at the bottom of the growth crucible (left: bottom view, right: side view). 
Grooves reduce the distance between melt and measurement wire to 0.5\,mm.
An additional indentation allows to place a thermocouple close to the measurement wire in order to scale resistivity to absolute temperature.
}
\label{Tiegelbearbeitung}
\end{center}
\end{figure} 

The thin tungsten measurement wire has two loose ends at each side of the crucible.  
These ends are connected to two tungsten rods (3\,mm diameter) that are lead through the cover plate similar to T1 reaching into the heating zone. 
The wire ends are connected to the rods by clamping them between the planar rod tip and a screw head fixed by a hex nut from the other parallel side. 
The electrical contact is stable up to at least $1000^{\circ}$C.
%The wire used for a resistivity measurement is connected to two tungsten rods (3\,mm diameter) that are lead through the coverplate identically to T1 reaching into the heating zone.
%The wire is connected to the rods by clamping them between the planar rod tip and a screw head fixed by a hex nut from the other parallel side. 
%The electrical contact is stable up to at least \mbox{1000$^{\circ}$C}.\\ 

Outside the furnace the rods are connected via coaxial cabels to a lock-in amplifier.
The resistivity at room temperature is given by measurement wire (\mbox{R $\thickapprox$ 7\,$\Omega$}), tungsten rods (\mbox{R $\thickapprox$ 5 m$\Omega$}) and coaxial cabel (\mbox{R $\thickapprox$ 70 m$\Omega$}). 
Accordingly the total resitivity is dominated by the voltage drop at the measurement wire (which is referred to as 'lock-in voltage' or $U(t)$ in the following).

\section{Experimental}\label{exp}

Investigated alloys were \textbf{Sb-In} (Sb: MaTeck, 99.9999\%; In: Chempur, 99.999\%), \textbf{Au-In} (Au: Edelmet. Recycling, 99.99\%), \textbf{Pd-In} (Pd: 99.95\%, Agosi), \textbf{Ni-Bi} (Ni: 99.98\%, GoodFellow; Bi: 99.9999\%, ChemPur), \textbf{Mn-Bi} (Mn: 99.99\%, MaTeck) and \textbf{Pd-Bi} (Pd: 99.95\%, Agosi; Bi: 99.999\%, 5NPlus).
\\
The used crucibles were made of aluminuma (LSP Industrial Ceramics Inc., diameter 12\,mm, height 25\,mm).
The grooves for the W wire were prepared at the \textit{Fraunhofer-Institut für Solare Energiesysteme ISE}, \textit{Wavering group} in Freiberg (Germany) using a high precision multi wire diamond saw.
The tungsten rods were obtained from ChemPur (99.95\%), the tungsten measurement wires from GoodFellow (99.95\%). 
The wires were fixed at the crucible by an aluminuma-based glue (Polytec GmbH, 903 HP, \mbox{$\upkappa$ = 5.7 $\frac{\text{W}}{\text{m} \cdot \text{K}}$}).\\
For measuring the voltage drop at the contact wire, the lock-in models \textit{7280 DSP} (Perkin \& Elmer; data shown in Figs.\,\ref{Alle_Testlegierungen}a-c,\,\ref{Verduennungsversuche}a-b) and \textit{MFLI 500kHz 60 MSa/s} (Zurich Instruments; data shown in Figs.\,\ref{SbIn},\,\ref{DTA-Vergleich_SbIn},\,\ref{Alle_Testlegierungen}d-f,\,\ref{AuIn_PdIn},\,\ref{Verduennungsversuche}c,\,\ref{PdBi}) were used. 
\\
DTA measurements of binary SbIn and PdBi (37:63) alloys were performed using a Netzsch \textit{STA 449 C/3/F Jupiter} DTA. 
An empty crucible was used as reference for SbIn, whereas the reference crucible of the PdBi sample was filled with bismuth of the same mass.
\\
Phase formation was verified by X-Ray powder diffraction using a \textit{Miniflex 600} diffractometer (Rigaku).

\section{Detecting Liquidus and solidus temperature of 'simple' binary alloys}\label{bina}

\begin{figure}
\includegraphics[width=0.8\columnwidth]{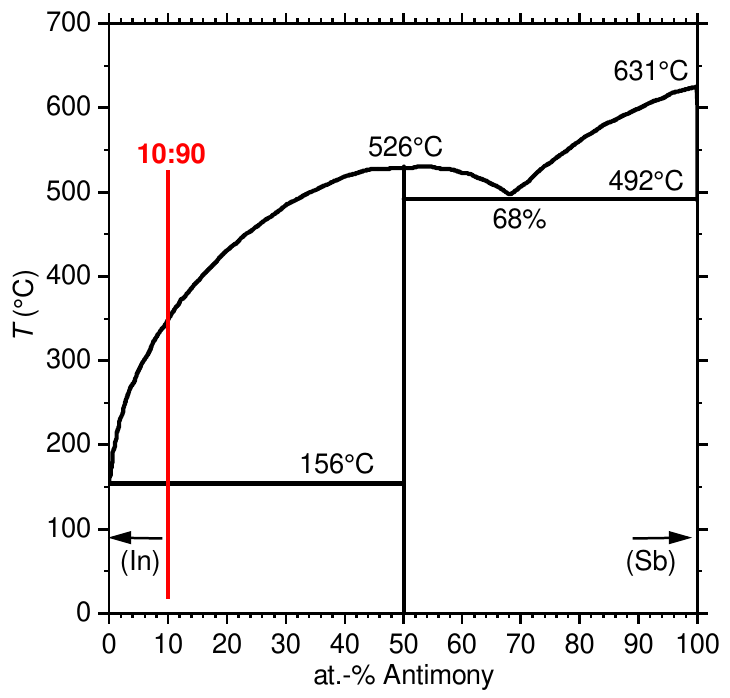}     
\caption{Binary alloy phase diagram of SbIn after Sharma \textit{et al.}\,\cite{Indium_Antimon_Phasendiagramm}. The composition of the starting materials is indicated by the vertical, red line.}
\label{Phasendiagramm_SbIn}
\end{figure}

Growth process, data collection and analysis are presented in detail for Sb-In. 
Lumps of antimony and indium were mixed in a molar ratio of Sb:In = 1:9.
According to the binary alloy phase diagram\,\cite{Indium_Antimon_Phasendiagramm}, two well separated phase transitions (at liquidus and solidus temperature) are expected for this composition (Fig.\,\ref{Phasendiagramm_SbIn}). 
The sample was heated up to 400$^{\circ}$C and kept there for 1\,h in order to homogenize the melt. 
Afterwards the heating was turned off and the solution allowed to cool down without further external influence.

\begin{figure}
\begin{center}
\includegraphics[width=0.8\columnwidth]{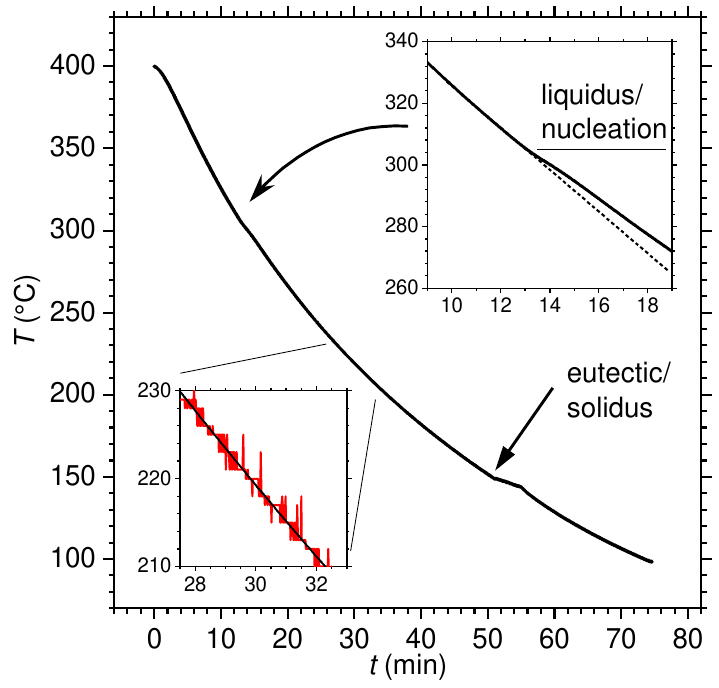}
\caption{Cooling process of an Sb$_{10}$In$_{90}$ alloy. 
The temperature profile was obtained from the lock-in voltage of the measurement wire. The corresponding values of the thermocouple temperature (T1, lower inset, red line) show identical time dependence with significantly lower resolution and were used to convert lock-in voltage to temperature.
Significant anomalies emerge at liquidus (inset) and solidus temperature.
}
\label{SbIn}
\end{center}
\end{figure}

The sample temperature as a function of time, $T1(t)$, is continuously measured by thermocouple T1. 
Changes in temperature cause a change of the electrical resistivity of the tungsten measurement wire.
This change in resistivity is directly proportional to the voltage-drop, $U(t)$, measured by the lock-in amplifier.
With time as an implicit parameter, $U(t)$ can be directly related to $T1(t)$. 
A plot of $U$ as a function of $T1$ reveals a basically linear dependence as expected for a metal.
Fitting a 2nd order polynomial to account for small deviations from linearity, yields an analytic expression for the temperature of the measurement wire $T(t) = f(U(t))$.
%With the data of T1 the voltage was converted to temperature. 
The time-dependent temperature $T(t)$ obtained for the Sb-In alloy is shown in Fig.\,\ref{SbIn}: the solidus is apparent and also the liquidus is observable with the naked eye (see upper inset).
The thermocouple readings T1 offer a much lower resolution and do not allow for a reliable detection of the phase transitions (lower inset in Fig.\,\ref{SbIn}).

\begin{figure}
\begin{center}
\includegraphics[width=0.85\columnwidth]{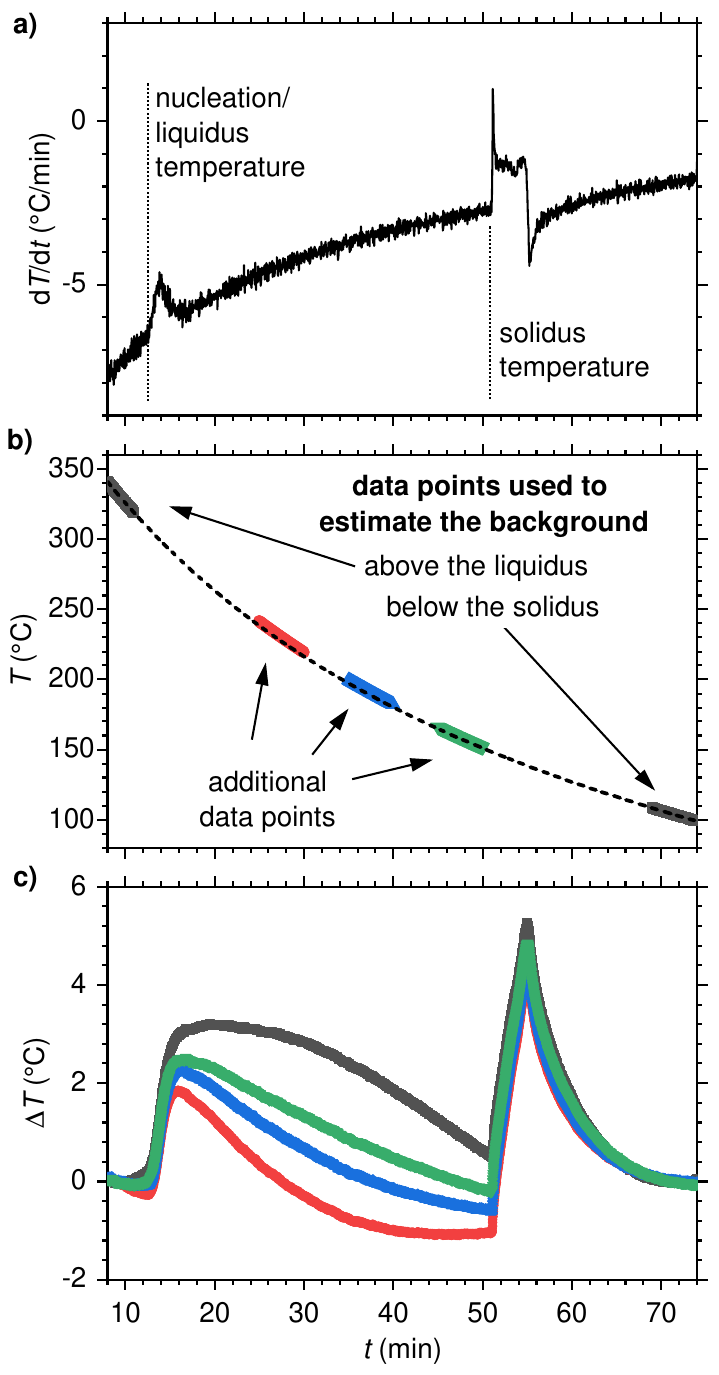}
\caption{Background determination $T_{\rm BG}(t)$, which reflects time-dependence of temperature unrelated to phase transitions.
a) derivative of the temperature $T(t)$ for an Sb$_{10}$In$_{90}$ alloy (see Fig.\,\ref{SbIn}).
b) interpolation of of $T(t)$ from above and below the phase transition temperatures yields $T_{\rm BG}(t)$ (dashed line). 
c) difference curves $\Delta T(t) = T(t) - T_{\rm BG}(t)$ obtained with and without  considering additional data points in between liquidus and solidus temperature in order to fit $T_{\rm BG}(t)$, from top to bottom: no additional data points, $t = 45-50$\,min, $t = 35-40$\,min, and $t = 25-30$\,min.}
\label{bg}
\end{center}
\end{figure}

The anomalies emerging at phase transitions are better observable in the difference curves $\Delta T(t) = T(t) - T_{\rm BG}(t)$ where $T_{\rm BG}(t)$ describes the time-dependence of temperature that is not related to phase transitions.
Additional measurements on empty crucibles confirmed the monotonic and featureless behavior of the background contribution.
The following steps present one possibility to determine $T_{\rm BG}(t)$ (see Fig\,\ref{bg}a-c):

a) the derivative d$T(t)$/d$t$ allows to estimate liquidus and solidus temperature (as well as possible other phase transitions).

b) part of the data set above liquidus and below solidus temperature are selected that show a time-dependence that is not affected by phase transitions (here $t = 5-11$\,min and $t = 69-74$\,min, see data points in dark grey).
Those data points were fitted to the following (empirical) function:
\begin{equation*}
T(t) = {\rm a} t + {\rm b} t^2 + T_{\rm eq} + (T_0 - T_{\rm eq})\,{\rm exp}(-t/\tau), 
\end{equation*}
with five free parameters $a, b, T_{\rm eq}, T_0,$ and $\tau$.
The obtained curve is shown by the black, dashed line.
Additional data points can be considered that are not affected by nucleation and solidification (see colored data points in regions $t = 25-30$\,min, $t = 35-40$\,min, and $t = 45-50$\,min for example).

c) The phase transition temperatures appear as sharp anomalies in $\Delta T(t)$ and do not depend on whether or how additional data points were used to fit $T_{\rm BG}(t)$. 
Since the details of $\Delta T(t)$ in the region between liquidus and solidus depend on the amount of latent heat released during the growth of SbIn, additional data points are required for higher precision. 
Given that $\Delta T(t)$ is supposed to be positive, the most reliable estimate is obtained by incorporating the region closest to the solidus (green data points, in accordance with the increasing slope of the liquidus line, see Fig.\,\ref{Phasendiagramm_SbIn}).
 
We want to emphasize that the presented determination and subtraction of $T_{\rm BG}(t)$ was primarily performed in order to allow for a direct comparison of feedback furnace results and DTA measurements (see below).
It is not necessary for an optimization of the growth procedure, which can be done by directly using the lock-in voltage $U(t)$ for the detection of nucleation as well as for controlling the heating power of the furnace.

\begin{figure}
\begin{center}
\includegraphics[width=0.85\columnwidth]{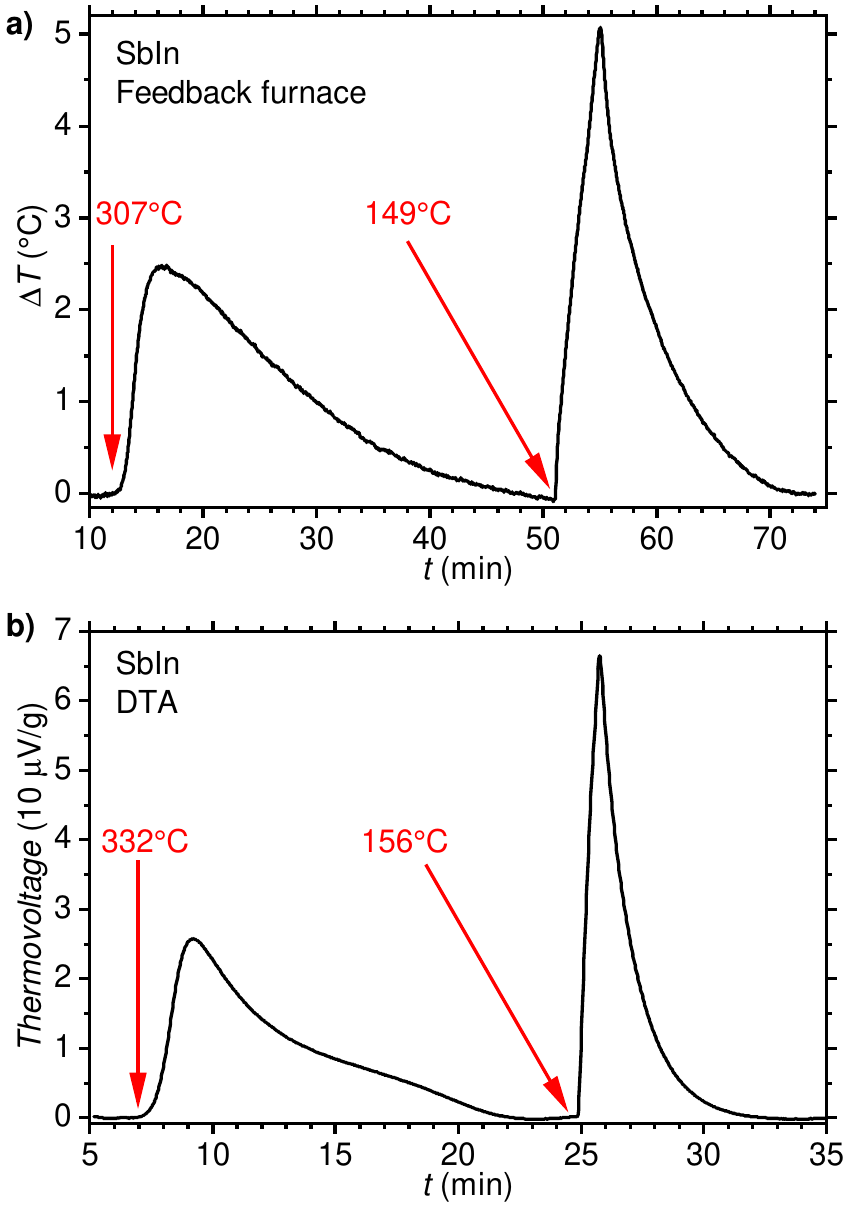}
\caption{Comparison of an Sb$_{10}$In$_{90}$ alloy cooled in the feedback furnace with a conventional DTA measurement. 
a) temperature profile obtained by the measurement wire (see Fig.\,\ref{SbIn}) after background subtraction.  
b) time-dependent thermovoltage measured by DTA.} 
\label{DTA-Vergleich_SbIn}
\end{center}
\end{figure}

Fig.\,\ref{DTA-Vergleich_SbIn}a shows $\Delta T(t)$ obtained for the Sb-In alloy ( $T_{\rm BG}(t)$ determined from above liquidus, below solidus and $t = 45-50$\,min).
In order to compare the signatures observed in $T(t)$, a DTA measurement was performed in a similar fashion (Fig.\,\ref{DTA-Vergleich_SbIn}b).
There is good agreement in the shape of characteristic signatures and the resolution of the different measurement techniques are comparable. 
In the feedback furnace, liquidus and solidus temperatures are determined to 307$^{\circ}$C and 149$^{\circ}$C, respectively. 
The respective temperatures measured in the DTA-system are 332$^{\circ}$C and 156$^{\circ}$C.
The liquidus temperatures differ due to different degrees of supercooling and are in reasonable agreement with previously reported values of 315$^{\circ}$C\,\cite{SbIn-Phasendiagramm-Abweichung_bei_10_Prozent} and $\sim 340^{\circ}$C\,\citep{Indium_Antimon_Phasendiagramm}.

\begin{figure}
\begin{center}
\includegraphics[width=\columnwidth]{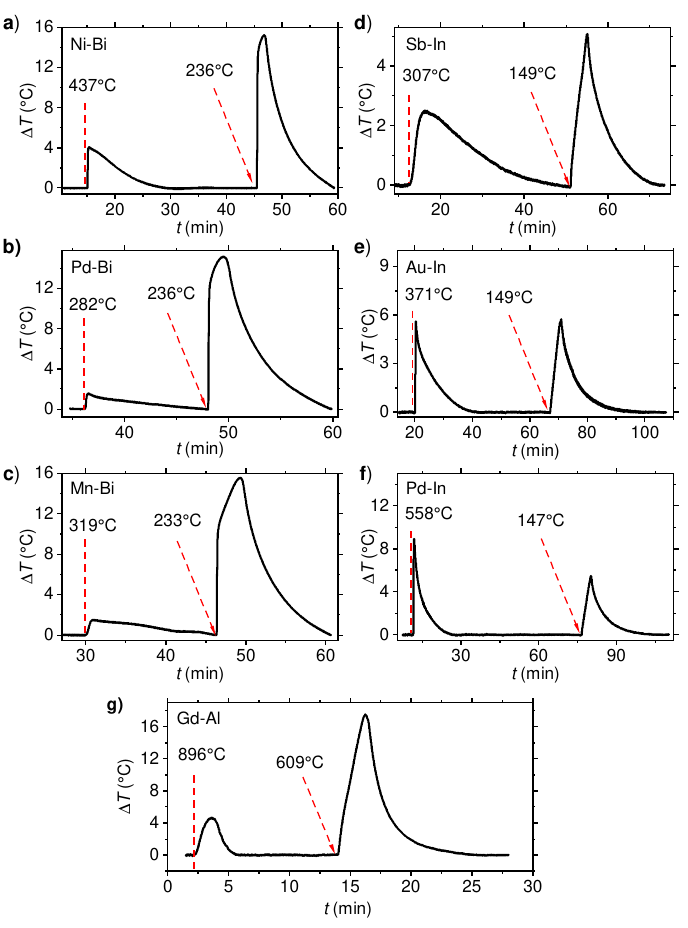}
\caption{Temperature profiles after background subtraction collected for several binary alloys (transition metal:Bi/In = 1:9). \textbf{a)} Ni-Bi, \textbf{b)} Pd-Bi, \textbf{c)}  Mn-Bi , \textbf{d)} Sb-In, \textbf{e)} Au-In, \textbf{f)} Pd-In, and \textbf{g)} Gd-Al 
(Temperature converted from lock-in voltage by the readings of thermocouple T1).}
\label{Alle_Testlegierungen}
\end{center}
\end{figure}

X-Ray powder diffraction was performed on ground SbIn single crystals. 
Those were separated from the flux by heating the solution above the eutectic and centrifuging afterwards\,\cite{Zentrifugieren_nach_Canfield}. 
The formation of the binary SbIn and small residuals of In-rich flux is confirmed.
\\
Comparable results were achieved with the six other binary systems Ni-Bi $\rightarrow$ NiBi$_3$ \cite{NiBi-Phasendiagramm}, 
Pd-Bi $\rightarrow$ PdBi$_2$ \cite{Palladium_Bismuth_Phasendiagramm}, 
Mn-Bi $\rightarrow$ MnBi \cite{MnBi_Phasendiagramm}, 
Au-In $\rightarrow$ AuIn$_2$ \cite{Gold_Indium_Phasendiagramm},
Pd-In $\rightarrow$PdIn$_3$ \cite{PdIn_Phasendiagramm}, and
Gd-Al $\rightarrow$AlGd$_2$ \cite{Massalski1986}
(the compound denoted after $\rightarrow$ crystallizes upon cooling below the liquidus temperature). 
The starting materials were mixed in the ratio transition metal/Gd:Bi/In/Al = 10:90.
The solutions of Ni-Bi, Mn-Bi, Au-In and Pd-Bi were homogenized for 60 minutes at 600$^{\circ}$C, the solution of Pd-In at 800$^{\circ}$C, and the solution of Gd-Al at 1000$^{\circ}$C.
Fig.\,\ref{Alle_Testlegierungen} shows the obtained signatures at liquidus and solidus temperatures after background subtraction. 
Significant variations are observed in the shape of the anomaly that is associated with nucleation: $\Delta T(t)$ increases step-like for Ni-Bi and Pd-Bi at $T = 437^{\circ}$ and $T = 282^{\circ}$, respectively (Fig.\ref{Alle_Testlegierungen}a,b) and even the overall slope of $T(t)$ (not shown) is positive indicating strong supercooling. 
This behavior was reproduced in four additional cooling procedures performed for both compounds. 
Mn-Bi and Sb-In (Fig.\ref{Alle_Testlegierungen}c,d), on the other hand, show a comparatively smooth increase in $\Delta T(t)$ at the liquidus temperature in all five measurements performed. 
The absence or at least a significantly lower tendency for supercooling is inferred.

\begin{figure}
\begin{center}
\includegraphics[width=\columnwidth]{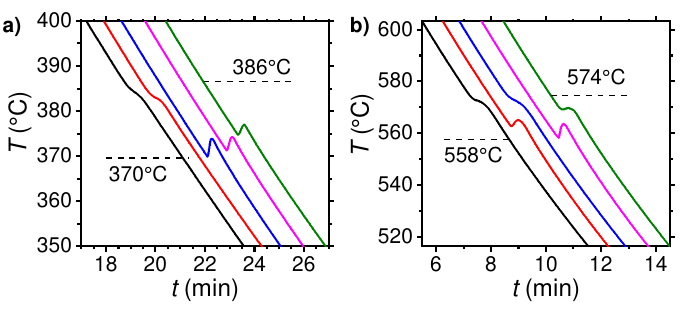}
\caption{Anomalies in the vicinity of the liquidus temperature recorded by the measurement wire in five consecutive runs (for clarity, curves are shifted horizontally keeping the order of measurements). 
a) mixture of Au:In = 1:9 with (run 1,2) and without (run 3,4,5) signatures of supercooling (indicated by sharp upturn in temperature).
b) mixture of Pd:In = 1:9 with alternating absence and occurrence of supercooling.
}
\label{AuIn_PdIn}
\end{center}
\end{figure}

Rather unexpected observations were made for the Au-In and Pd-In alloys, with strong supercooling shown in Fig.\ref{Alle_Testlegierungen}e,f: 
successive cooling procedures indicate strong supercooling in some cases whereas no such signatures are observed in other runs. 
This is best seen in the raw data prior to background subtraction (Fig.\,\ref{AuIn_PdIn}, the temperature $T(t)$ is obtained by converting the lock-in voltage).
Note that those runs were performed consecutively without adjusting the measurement technique or opening the growth chamber.
In the Au-In cooling curves 1 and 2 (Fig.\ref{AuIn_PdIn}a), a change of slope emerges at $T = 386^{\circ}$C and marks the liquidus temperature.
In runs 3, 4, and 5, on the other hand, the temperature increases in a step-like fashion at significantly lower temperature of $T = 370^\circ$C.
Both, shape of the anomaly and reduced temperature, indicate supercooling (by roughly $16^\circ$C).
Even alternating behavior is observed for Pd-In (Fig.\ref{AuIn_PdIn}b). 
A relation between characteristic temperature and shape of the anomaly is recognizable: strong supercooling in runs 2 and 4 as evidenced by the lowest characteristic temperature and a strong increase in $T(t)$ at the nucleation.
Runs 1 and 3 show the anomaly at significantly higher temperature ($+16^{\circ}$C) and no increase in $T(t)$.
Intermediate behavior is observed in run 5 with a moderate increase in temperature at intermediate characteristic temperature.

So far, the presented phase transitions took place at rather low temperature, with events all below $T = 600^{\circ}$C. As shown in Fig.\,\ref{Alle_Testlegierungen}g for an Al-rich solution, successful detection is possible at significantly higher temperature of $T \approx 900^{\circ}$C.

\section{Resolution limit}\label{dilute}

\begin{figure}
\begin{center}
\includegraphics[width=0.85\columnwidth]{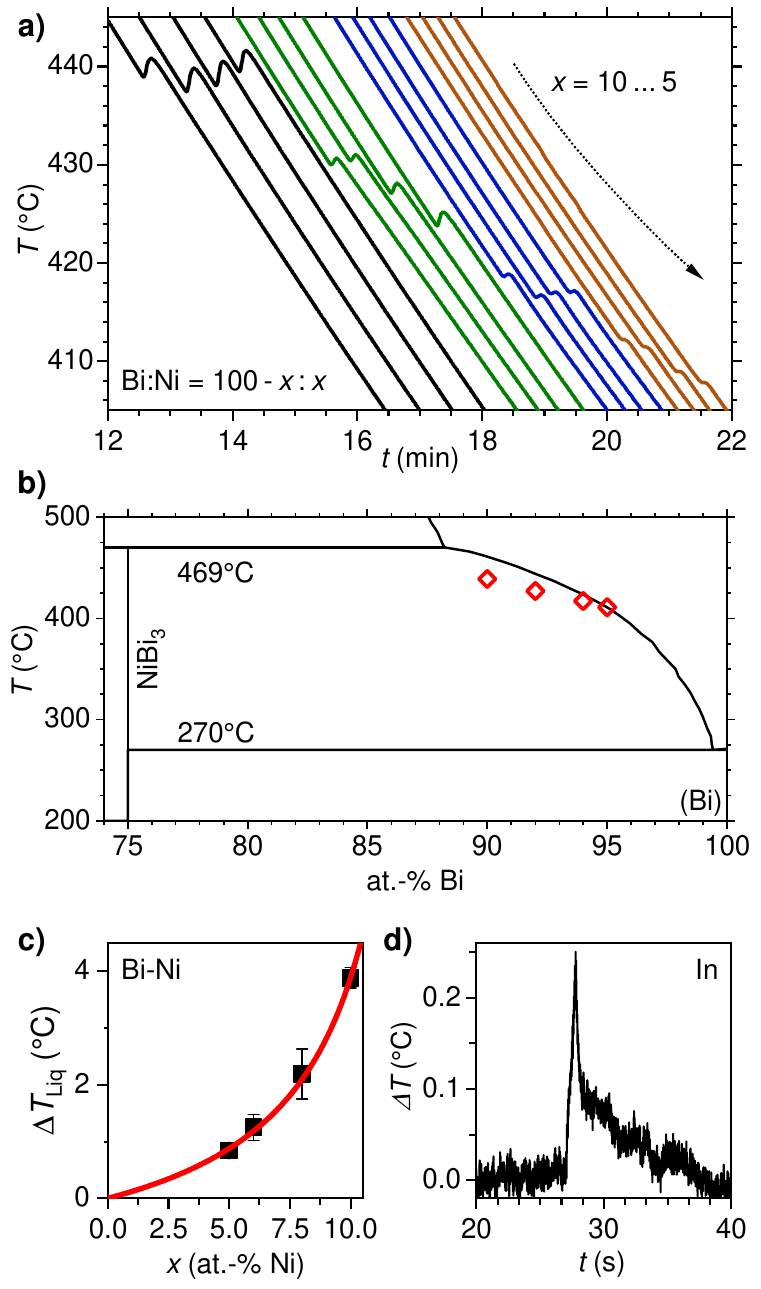}
\caption{Resolution limit of the feedback furnace a) Temperature profile recorded by the measurement wire for a series of Ni-Bi alloys mixed in ratios Ni:Bi = 10:90 (black curves), 8:92 (green curves), 6:94 (blue curves), and 5:95 (brown curves).
The expected decrease of the liquidus temperature for increasing Bi concentration\,\cite{NiBi-Phasendiagramm} is apparent (for clarity, curves are shifted horizontally keeping the order of measurements). 
b) Ni-Bi binary phase diagram expanded for the Bi-rich side\,\cite{NiBi-Phasendiagramm} with liquidus temperatures obtained in the feedback furnace shown by open, red symbols.
c) Magnitude of the anomaly in $\Delta T(t)$ at the liquidus temperature as a function of the Ni concentration (error bar represents the variation over the four runs, the line is a fit to a hyperbolic function - see text). 
d) Solidification of 46\,mg In at $T = 151^{\circ}$C.
}
\label{Verduennungsversuche}
\end{center}
\end{figure}

Dilution experiments were performed on a Bi-rich flux with the starting elements mixed in a ratio of Bi:Ni = 9:1.
After detecting the nucleation of NiBi$_3$ in several runs, the Bi content was increased by adding Bi pieces to the crucible, such that the Ni concentration decreased to Bi:Ni = 92:8, 94:6 and 95:5.
The obtained time-dependent temperatures at the crucible $T(t)$ (obtained by converting the lock-in voltage) are shown in Fig.\,\ref{Verduennungsversuche}a.
As expected from the binary alloy phase diagram\,\cite{NiBi-Phasendiagramm}, the liquidus temperature decreases with increasing Bi concentration. 
Furthermore, the anomaly appears weaker due to the smaller amount of precipitating material.
With the data at hand, we cannot unambiguously tell whether the tendency to supercooling is also reduced or an increase in $T(t)$ at the nucleation is unobservable due to the smaller amount of latent heat.
A comparison of obtained liquidus temperatures with the binary alloy phase diagram\,\cite{NiBi-Phasendiagramm} indicates weaker supercooling for lower Ni content (Fig.\,\ref{Verduennungsversuche}b).

In order to estimate the lowest Ni concentration that still allows for the detection of NiBi$_3$ nucleation, we analyzed the dependence of the anomaly in $\Delta T(t)$ as a function of the Ni concentration (after background subtraction).
The obtained step height at the anomaly, $\Delta T_{\rm Liq}$, is shown in Fig.\,\ref{Verduennungsversuche}c.
The error bar represents the variation over the four runs (standard deviation).
The solid line is a fit to to a hyperbolic function [$\Delta T_{\rm Liq} = \alpha x/(\beta-x)$] that describes the expected decrease of the anomaly with decreasing Ni concentration.  
Given that relative changes in temperature (obtained from the lock-in voltage) are detectable on a level of $10^{-4}$, the resolution limit is reached for an estimated Ni concentration of roughly 0.25\,at.-\%.
\\ 
In addition, the resolution limit was investigated by measuring a small piece of In ($m = 45.8$\,mg).
A clear signature in $\Delta T(t)$ is observed at the solidification at $T = 151^{\circ}$C\,\ (Fig.\,\ref{Verduennungsversuche}d) in good agreement with the melting temperature of $T = 156^{\circ}$C\,\cite{Handbook}.
The step in $\Delta T(t)$ amounts to 0.26$^{\circ}$C and is roughly 10 times larger than the noise level. 
With the enthalpy of fusion of In, $\Delta_{\rm fus}H = 28.6$\,J/g\,\cite{Archer2003}, we estimate that phase transitions with latent heat release of $\gtrsim 100$\,mJ are detectable with the naked eye. 

\section{Results on P\MakeLowercase{d}$_{37}$B\MakeLowercase{i}$_{63}$}\label{pdbi}

\begin{figure}
\begin{center}
\includegraphics[width=0.9\columnwidth]{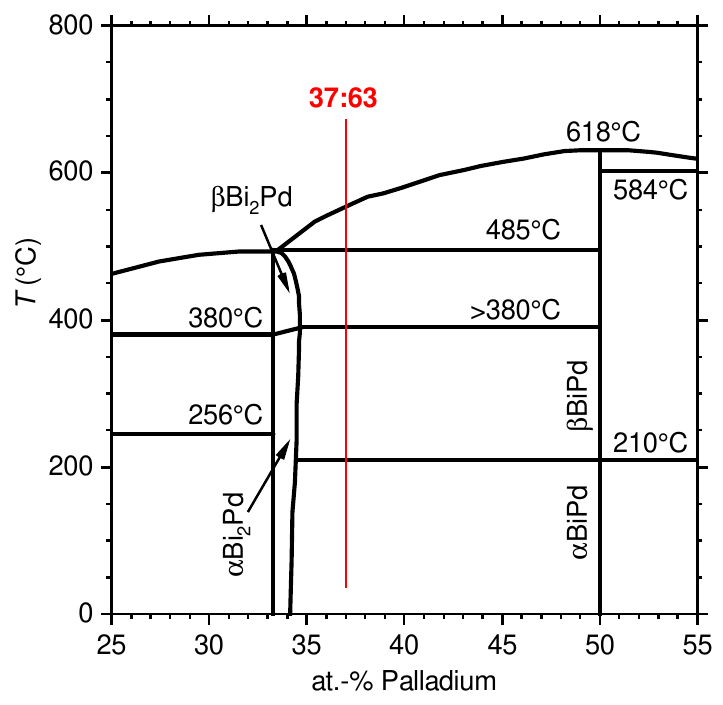}
\caption{Binary alloy phase diagram of PdBi\,\cite{Palladium_Bismuth_Phasendiagramm, Bhatt1979}.
The composition of the starting materials is indicated by the vertical, red line.}
\label{PdBi_Phasendiagramm}
\end{center}
\end{figure}

A more complex series of phase transitions is expected for PdBi mixed in a ratio of Pd:Bi = 37:63 as inferred from the binary alloy phase diagram shown in Fig.\,\ref{PdBi_Phasendiagramm}\,\citep{Palladium_Bismuth_Phasendiagramm}.
Even though the target compound PdBi melts congruently and can be grown by the Bridgman technique\,\cite{Okawa2013, Peets2014, Sun2015}, flux growth\,\cite{Thirupathaiah2016} offers several advantages: free growth of faceted crystals in the natural habit, 'in-situ purification' of the starting materials, (potentially) higher diffusion rates and the use of comparatively small amounts of material.
We have chosen PdBi primarily because it offers a phase diagram of intermediate complexity and structural phase transitions in addition to liquidus and solidus. 
%The crucible was loosely covered by an aluminuma cap with a piece of Zr-foil ontop as oxygen getter.

In contrast to the previous processes, regular heat pulses of 1200 ms every 10 seconds were used for heating instead of controlling the temperature by a setpoint. 
This resulted in a temperature profile that is not bound to a predetermined slope and, more importantly, that does not mask the anomalies caused by phase transitions.
Large heating rates are employed and the maximum temperature of 700$^{\circ}$C is reached in roughly 27 minutes. 
Accordingly, the absorption of latent heat takes place over comparatively small periods in time and the associated anomalies in $T(t)$ are better observable.
Note that furnace cooling requires roughly 1\,h for reducing temperature from $700^{\circ}$C to $200^{\circ}$C.

\begin{figure}
\begin{center}
\includegraphics[width=0.9\columnwidth]{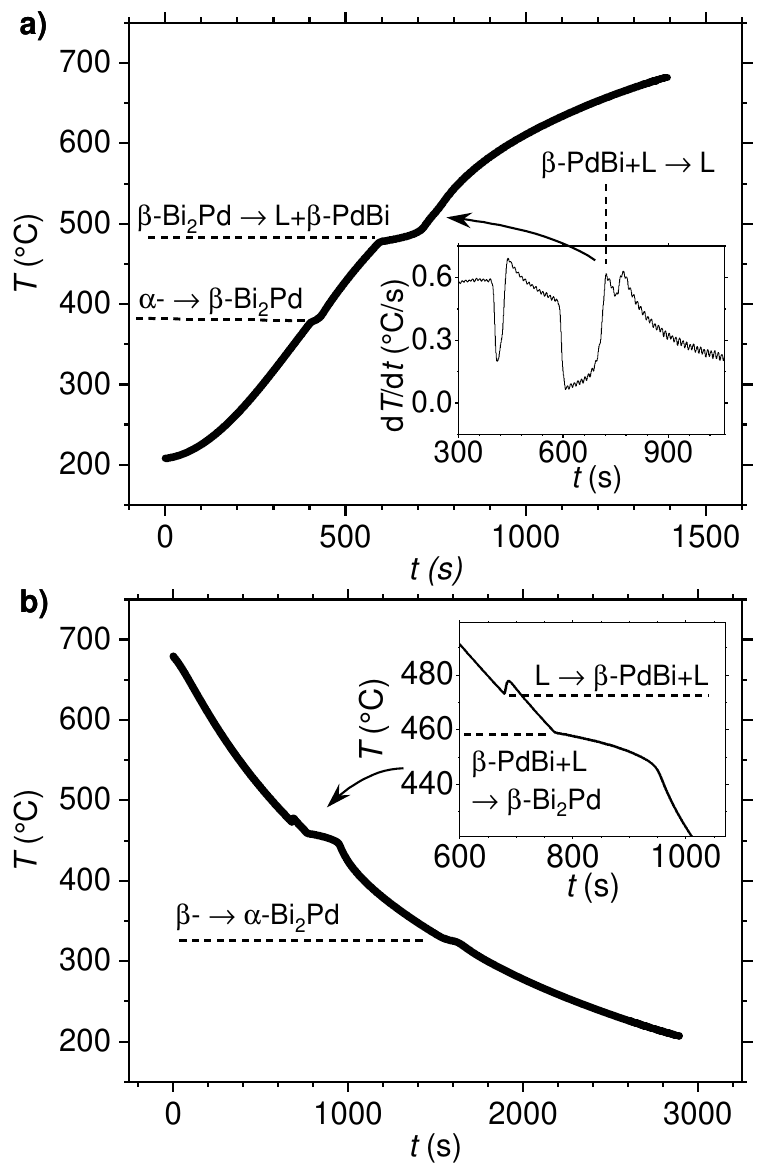}
\caption{Temperature profile of Pd:Bi = 37:63 recorded in the feedback furnace. 
a) Structural phase transition ($\alpha \rightarrow \beta$ Bi$_2$Pd) and melting of Bi$_2$Pd manifest in well defined kinks in $T(t)$. The liquidus temperature is observable in the derivative d$T(t)$/d$t$ (inset).
b) significant supercooling is indicated by the sharp increase in $T(t)$ (inset) and the lowering of the transition temperature by $\sim40^\circ$ when compared to the heating curve.
Solidification and structural phase transition of Bi$_2$Pd are well resolved.
}
\label{PdBi}
\end{center}
\end{figure}
 
Fig.\,\ref{PdBi}a,b shows the temperature profiles obtained in the feedback furnace.
All phase transitions have been assigned according to the PdBi binary phase diagram \cite{Palladium_Bismuth_Phasendiagramm}. 
%One of those phase transitions is assigned to a structural phase transition. 
The extracted transition temperatures are shown in table\,\ref{Übergangstemperaturen} together with the results obtained by conventional DTA (see below) and literature data\,\cite{Palladium_Bismuth_Phasendiagramm}.
Phase formation was confirmed by XRD measurements on ground single crystals.
Flux removal was performed by decanting above the melting point of Bi$_2$Pd using a frit-disc crucible\,\cite{Zentrifugieren_nach_Canfield}.
Residual flux was removed by polishing the crystals before the grinding. 
The diffraction pattern showed the presence of $\alpha$-PdBi as main phase with traces of PdBi$_2$.

\begin{table}
\begin{center}
\begin{tabular}{c c c c c c}
\hline
\hline
\textbf{Transition}&\textbf{Lit.}\cite{Palladium_Bismuth_Phasendiagramm}&\textbf{Furnace$\uparrow$}&\textbf{DTA$\uparrow$}&\textbf{Furnace$\downarrow$} & \textbf{DTA$\downarrow$}\\	
\hline															 
$\upalpha$ $\leftrightarrow$ $\upbeta$-PdBi$_2$& >380$^{\circ}$C& \textcolor{red}{376$^{\circ}$C} &\textcolor{red}{378$^{\circ}$C}&\textcolor{blue}{333$^{\circ}$C} & \textcolor{blue}{328$^{\circ}$C} \\
$\upbeta$ -PdBi$_2$ $\leftrightarrow$ L&485$^{\circ}$C & \textcolor{red}{476$^{\circ}$C} &\textcolor{red}{485$^{\circ}$C}& \textcolor{blue}{458$^{\circ}$C}  &  \textcolor{blue}{474$^{\circ}$C}\\
$\upbeta$-PdBi $\leftrightarrow$ L&520$^{\circ}$C & \textcolor{red}{510$^{\circ}$C} &\textcolor{red}{522$^{\circ}$C}& \textcolor{blue}{473$^{\circ}$C}  & \textcolor{blue}{478$^{\circ}$C}\\
\hline
\hline
\end{tabular}
\caption{Detected and assigned phase transitions of Pd-Bi (37:63) acquired by feedback furnace and DTA (heating:$\uparrow$, cooling:$\downarrow$)}
\label{Übergangstemperaturen}
\end{center}
\end{table}

\begin{figure}
\begin{center}
\includegraphics[width=\columnwidth]{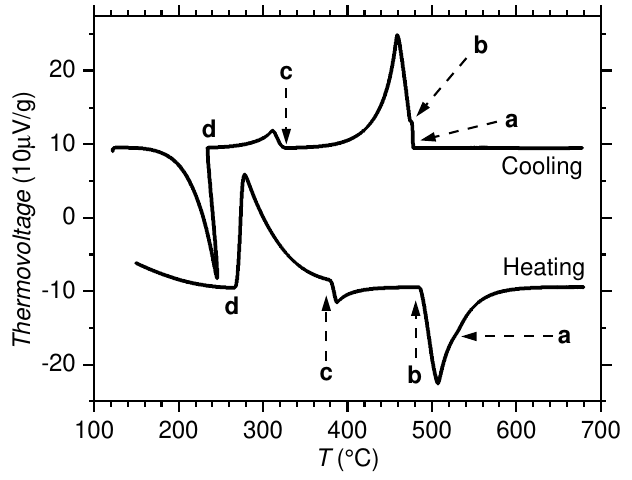}
\caption{DTA measurement of Pd:Bi = 37:63 performed with heating and cooling rates of $20^\circ$C/min. Phase transition are assigned according to the binary alloy phase diagram. \textbf{a}: liquidus temperature, \textbf{b}: Melting and solidification of $\upbeta$-PdBi$_2$, \textbf{c}: Structural phase transition PdBi$_2$ and \textbf{d}: melting and solidification of Bi in the reference crucible}
\label{PdBi_DTA}
\end{center}
\end{figure}

The values of the detected phase transitions get fortified by DTA measurements presented in Fig.\,\ref{PdBi_DTA}. 
The reference crucible was filled with elemental bismuth that caused additional anomalies  during melting and solidification. 
The cooling rate of $20^\circ$C/min was similar to the one present in the feedback furnace prior to the liquidus temperature. 
The reported structural phase transition from $\beta$-PdBi to $\alpha$-PdBi at $T = 210^\circ$ \cite{Bhatt1979} remains elusive in DTA and feedback furnace. 

\section{Discussion}\label{disc}
In the following, we discuss the advantages and development potential of the feedback furnace by elaborating on the difficulties associated with the standard procedure of collecting and using information on phase transitions prior to the growth process in standard laboratory furnaces.
DTA and the related DSC (Differential scanning calorimetry) are well established techniques for detecting phase transitions in various kinds of condensed matter.
Starting from the 1940s\,\cite{DTA_Geschichte}, methods and implementations were constantly improved\,\cite{Chiu2011}.
The application of DTA and DSC results to solution growth is, however, not straightforward and connected with various challenges.
First of all, the actual spread of determined phase transition temperatures given in phase diagrams can be larger than anticipated from the (normally) smooth lines plotted; see, for example, a detailed study on the Li-N system published by Sangster and Pelton\,\cite{Sangster1992} and compare with the standard presentation shown in the renowned phase diagram collections\,\cite{Okamoto1998_LiN}. 
Various further problems are encountered in the application of DTA to crystallization from high-temperature solutions in particular for reactive materials\,\cite{Elwell1969, Neate1971}.

Furthermore, it is well known that the actual temperature of the melt can significantly deviate from the nominal temperature displayed by standard laboratory furnaces. 
Accordingly, even highly accurate DTA/DSC results are not always applicable with nominal precision in particular when aging of thermocouples and temperature-dependence of gradients are taken into account.
We avoid these problems by utilizing the feedback method: The hitherto crucial measurement of the absolute temperature merely serves as an additional source of information. The state of a sample is determined not by the nominal temperature in the process chamber, but by the onset or absence of characteristic signatures directly originating in the crystallization processes of the sample material.

Supercooling presents another difficulty that hampers the application of knowledge on liquidus temperatures determined by DTA/DSC. 
The degree of supercooling can depend on purity of starting materials, crucible material and crucible shape.
The feedback furnace allows estimating the tendency for supercooling by the shape of nucleation anomalies and comparing heating and cooling curves directly in the crystal growth setup. 
This enables extremely slow cooling rates or extended hold times in a barely supercooled state.
Detecting the nucleation and estimating the degree of supercooling opens the possibility to perform 'seed selection' by oscillating the temperature around the liquidus.

%As for example the different temperatures for the transition \mbox{L $\rightarrow$ $\upbeta$-PdBi$_2$} of furnace and DTA are showing, even samples equal in ratio may lead to a different result, 15$^{\circ}$C in this case. 
%Especially if the solidification of the desired compound (L$\rightarrow$ $\upbeta$-PdBi) is very close to the next transition (L$\rightarrow$ $\upbeta$-PdBi$_2$), even a single degree of inaccuracy may become crucial. 
%
% 
%The same applies for the transition of \mbox{$\upbeta$ -PdBi$_2$} $\rightarrow$ L, which is 485$^{\circ}$C in the DTA and 480$^{\circ}$C in the furnace (s. Tab. \ref{Übergangstemperaturen}).
%Still other temperature values, such as the phase transition of $\upalpha$-PdBi$_2$ $\rightarrow$ $\upbeta$-PdBi$_2$ (382$^{\circ}$C $\leftrightarrow$ >380$^{\circ}$C \cite{Palladium_Bismuth_Phasendiagramm}) are in good agreement with the respective literature. Regardless of any slight deviations in the assigned temperature, the state of the system is determined by the absence or occurence of an anomaly, not by its abolute temperature.\\ 
%

\section{Summary}\label{sum}
We present a novel phase transition detection method and its implementation in a crystal growth setup for solution growth in controlled atmospheres at temperatures up to $1000^\circ$C.
A thin measurement wire is mounted in close vicinity to the melt and acts as highly precise, resistive thermometer.
Anomalies in the time-dependent temperature are associated with phase transitions in the sample material.   
In contrast to established DTA and DSC measurements, there is no reference crucible employed allowing for an efficient and symmetric design of the growth chamber.
The feedback furnace combines crystal growth with a thermal analysis of the sample material. Both heretofore locally and temporally separated processes take place in the same crucible at the same time, as the furnace detection method gets applied at the actual solution growth sample, showing the same anomalies without a necessary transfer from detection to growth apparatus. The furnace detection and growth system provides a method has that offers a considerable increase in efficiency and accuracy for many solution growth processes.
 
Liquidus and solidus temperature were successfully detected in seven different binary alloys in good agreement with DTA results.
An additional structural transition was detected in Bi$_2$Pd.
Shape of the characteristic anomalies and comparison of heating and cooling curves allow to estimate the degree of supercooling.
Given that there is no direct contact of measurement wire and melt, there are no restrictions on the choice of the (ceramic) crucible material beyond standard requirements.
An optimized version of the feedback furnace is going to be available\,\cite{scidrewebsite2}.
With the in-situ detection in the feedback furnace, a method has been established that offers a considerable increase in efficiency and accuracy for various solution growth processes.
Further improvements of sensitivity and precision by higher sampling rates and advanced, real-time data analysis are anticipated. 

\section{Acknowledgement}
We thank Alexander Herrnberger, Klaus Wiedenmann, Patrick Schütze, and Maik Jurischka for technical support. 
This work was supported by the Deutsche Forschungsgemeinschaft (DFG, German Research
Foundation) - Grant No. JE748/1 and the ZIM program of the German Ministry for Economic Affairs and Energy (BMWi). 

\bibliography{Quellenverzeichnis}

%merlin.mbs apsrev4-1.bst 2010-07-25 4.21a (PWD, AO, DPC) hacked
%Control: key (0)
%Control: author (8) initials jnrlst
%Control: editor formatted (1) identically to author
%Control: production of article title (-1) disabled
%Control: page (0) single
%Control: year (1) truncated
%Control: production of eprint (0) enabled
\begin{thebibliography}{29}%
\makeatletter
\providecommand \@ifxundefined [1]{%
 \@ifx{#1\undefined}
}%
\providecommand \@ifnum [1]{%
 \ifnum #1\expandafter \@firstoftwo
 \else \expandafter \@secondoftwo
 \fi
}%
\providecommand \@ifx [1]{%
 \ifx #1\expandafter \@firstoftwo
 \else \expandafter \@secondoftwo
 \fi
}%
\providecommand \natexlab [1]{#1}%
\providecommand \enquote  [1]{``#1''}%
\providecommand \bibnamefont  [1]{#1}%
\providecommand \bibfnamefont [1]{#1}%
\providecommand \citenamefont [1]{#1}%
\providecommand \href@noop [0]{\@secondoftwo}%
\providecommand \href [0]{\begingroup \@sanitize@url \@href}%
\providecommand \@href[1]{\@@startlink{#1}\@@href}%
\providecommand \@@href[1]{\endgroup#1\@@endlink}%
\providecommand \@sanitize@url [0]{\catcode `\\12\catcode `\$12\catcode
  `\&12\catcode `\#12\catcode `\^12\catcode `\_12\catcode `\%12\relax}%
\providecommand \@@startlink[1]{}%
\providecommand \@@endlink[0]{}%
\providecommand \url  [0]{\begingroup\@sanitize@url \@url }%
\providecommand \@url [1]{\endgroup\@href {#1}{\urlprefix }}%
\providecommand \urlprefix  [0]{URL }%
\providecommand \Eprint [0]{\href }%
\providecommand \doibase [0]{http://dx.doi.org/}%
\providecommand \selectlanguage [0]{\@gobble}%
\providecommand \bibinfo  [0]{\@secondoftwo}%
\providecommand \bibfield  [0]{\@secondoftwo}%
\providecommand \translation [1]{[#1]}%
\providecommand \BibitemOpen [0]{}%
\providecommand \bibitemStop [0]{}%
\providecommand \bibitemNoStop [0]{.\EOS\space}%
\providecommand \EOS [0]{\spacefactor3000\relax}%
\providecommand \BibitemShut  [1]{\csname bibitem#1\endcsname}%
\let\auto@bib@innerbib\@empty
%</preamble>
\bibitem [{\citenamefont {Elwell}\ and\ \citenamefont
  {Scheel}(1975)}]{Elwell1975}%
  \BibitemOpen
  \bibfield  {author} {\bibinfo {author} {\bibfnamefont {D.}~\bibnamefont
  {Elwell}}\ and\ \bibinfo {author} {\bibfnamefont {H.~J.}\ \bibnamefont
  {Scheel}},\ }\href@noop {} {\emph {\bibinfo {title} {Crystal Growth from
  High-Temperature Solutions}}}\ (\bibinfo  {publisher} {Academic Press Inc.,
  London},\ \bibinfo {year} {1975})\BibitemShut {NoStop}%
\bibitem [{\citenamefont {Canfield}\ and\ \citenamefont
  {Fisher}(2001)}]{Canfield2001}%
  \BibitemOpen
  \bibfield  {author} {\bibinfo {author} {\bibfnamefont {P.~C.}\ \bibnamefont
  {Canfield}}\ and\ \bibinfo {author} {\bibfnamefont {I.~R.}\ \bibnamefont
  {Fisher}},\ }\href {\doibase 10.1016/S0022-0248(01)00827-2} {\bibfield
  {journal} {\bibinfo  {journal} {J. Cryst. Growth}\ }\textbf {\bibinfo
  {volume} {225}},\ \bibinfo {pages} {155} (\bibinfo {year}
  {2001})}\BibitemShut {NoStop}%
\bibitem [{\citenamefont {Kanatzidis}\ \emph {et~al.}(2005)\citenamefont
  {Kanatzidis}, \citenamefont {P\"ottgen},\ and\ \citenamefont
  {Jeitschko}}]{Flusszuechtung}%
  \BibitemOpen
  \bibfield  {author} {\bibinfo {author} {\bibfnamefont {M.~M.}\ \bibnamefont
  {Kanatzidis}}, \bibinfo {author} {\bibfnamefont {R.}~\bibnamefont
  {P\"ottgen}}, \ and\ \bibinfo {author} {\bibfnamefont {W.}~\bibnamefont
  {Jeitschko}},\ }\href@noop {} {\bibfield  {journal} {\bibinfo  {journal}
  {Angew. Chem. Int. Ed., 44, 6996 - 7023}\ } (\bibinfo {year}
  {2005})}\BibitemShut {NoStop}%
\bibitem [{\citenamefont {Jesche}\ and\ \citenamefont
  {Canfield}(2014)}]{Jesche2014c}%
  \BibitemOpen
  \bibfield  {author} {\bibinfo {author} {\bibfnamefont {A.}~\bibnamefont
  {Jesche}}\ and\ \bibinfo {author} {\bibfnamefont {P.~C.}\ \bibnamefont
  {Canfield}},\ }\href {\doibase 10.1080/14786435.2014.913114} {\bibfield
  {journal} {\bibinfo  {journal} {Philos. Mag.}\ }\textbf {\bibinfo {volume}
  {94}},\ \bibinfo {pages} {2372} (\bibinfo {year} {2014})}\BibitemShut
  {NoStop}%
\bibitem [{\citenamefont {{https://scidre.de}}()}]{scidrewebsite}%
  \BibitemOpen
  \bibfield  {author} {\bibinfo {author} {\bibnamefont {{https://scidre.de}}},\
  }\href@noop {} {}\BibitemShut {NoStop}%
\bibitem [{\citenamefont {Fischer}(1966)}]{Fischer1966}%
  \BibitemOpen
  \bibfield  {author} {\bibinfo {author} {\bibfnamefont {A.~K.}\ \bibnamefont
  {Fischer}},\ }\href {\doibase 10.1063/1.1727337} {\bibfield  {journal}
  {\bibinfo  {journal} {J. Chem. Phys.}\ }\textbf {\bibinfo {volume} {45}},\
  \bibinfo {pages} {375} (\bibinfo {year} {1966})},\ \Eprint
  {http://arxiv.org/abs/https://doi.org/10.1063/1.1727337}
  {https://doi.org/10.1063/1.1727337} \BibitemShut {NoStop}%
\bibitem [{\citenamefont {Sharma}\ \emph {et~al.}(1989)\citenamefont {Sharma},
  \citenamefont {Ngai},\ and\ \citenamefont
  {Chang}}]{Indium_Antimon_Phasendiagramm}%
  \BibitemOpen
  \bibfield  {author} {\bibinfo {author} {\bibfnamefont {R.}~\bibnamefont
  {Sharma}}, \bibinfo {author} {\bibfnamefont {T.}~\bibnamefont {Ngai}}, \ and\
  \bibinfo {author} {\bibfnamefont {Y.}~\bibnamefont {Chang}},\ }\href@noop {}
  {\bibfield  {journal} {\bibinfo  {journal} {Bull Alloy Phase Diagr.}\ }
  (\bibinfo {year} {1989})}\BibitemShut {NoStop}%
\bibitem [{\citenamefont {Liu}\ and\ \citenamefont
  {Perreti}(1952)}]{SbIn-Phasendiagramm-Abweichung_bei_10_Prozent}%
  \BibitemOpen
  \bibfield  {author} {\bibinfo {author} {\bibfnamefont {T.}~\bibnamefont
  {Liu}}\ and\ \bibinfo {author} {\bibfnamefont {E.}~\bibnamefont {Perreti}},\
  }\href@noop {} {\bibfield  {journal} {\bibinfo  {journal} {Trans. ASM, 44,
  539-548}\ } (\bibinfo {year} {1952})}\BibitemShut {NoStop}%
\bibitem [{\citenamefont {Canfield}\ \emph {et~al.}(2016)\citenamefont
  {Canfield}, \citenamefont {Kong}, \citenamefont {Kaluarachchi},\ and\
  \citenamefont {Jo}}]{Zentrifugieren_nach_Canfield}%
  \BibitemOpen
  \bibfield  {author} {\bibinfo {author} {\bibfnamefont {P.}~\bibnamefont
  {Canfield}}, \bibinfo {author} {\bibfnamefont {T.}~\bibnamefont {Kong}},
  \bibinfo {author} {\bibfnamefont {U.}~\bibnamefont {Kaluarachchi}}, \ and\
  \bibinfo {author} {\bibfnamefont {N.~H.}\ \bibnamefont {Jo}},\ }\href@noop {}
  {\bibfield  {journal} {\bibinfo  {journal} {Philos. Mag., 96, 84-92}\ }
  (\bibinfo {year} {2016})}\BibitemShut {NoStop}%
\bibitem [{\citenamefont {Voss}(1908)}]{NiBi-Phasendiagramm}%
  \BibitemOpen
  \bibfield  {author} {\bibinfo {author} {\bibfnamefont {G.}~\bibnamefont
  {Voss}},\ }\href@noop {} {\bibfield  {journal} {\bibinfo  {journal} {Z.
  Anorg. Allg. Chem. 57, 52-58}\ } (\bibinfo {year} {1908})}\BibitemShut
  {NoStop}%
\bibitem [{\citenamefont {Brasier}\ and\ \citenamefont
  {Hume-Rothery}(1959)}]{Palladium_Bismuth_Phasendiagramm}%
  \BibitemOpen
  \bibfield  {author} {\bibinfo {author} {\bibfnamefont {J.}~\bibnamefont
  {Brasier}}\ and\ \bibinfo {author} {\bibfnamefont {W.}~\bibnamefont
  {Hume-Rothery}},\ }\href@noop {} {\bibfield  {journal} {\bibinfo  {journal}
  {J. Less-Common Met. (157-164), Vol. 1}\ } (\bibinfo {year}
  {1959})}\BibitemShut {NoStop}%
\bibitem [{\citenamefont {Siebe}(1919)}]{MnBi_Phasendiagramm}%
  \BibitemOpen
  \bibfield  {author} {\bibinfo {author} {\bibfnamefont {P.}~\bibnamefont
  {Siebe}},\ }\href@noop {} {\bibfield  {journal} {\bibinfo  {journal} {Z.
  Anorg. Allg. Chem. 108, 161-171}\ } (\bibinfo {year} {1919})}\BibitemShut
  {NoStop}%
\bibitem [{\citenamefont {Hiscocks}\ and\ \citenamefont
  {Hume-Rothery}(1964)}]{Gold_Indium_Phasendiagramm}%
  \BibitemOpen
  \bibfield  {author} {\bibinfo {author} {\bibfnamefont {S.}~\bibnamefont
  {Hiscocks}}\ and\ \bibinfo {author} {\bibfnamefont {W.}~\bibnamefont
  {Hume-Rothery}},\ }\href@noop {} {\bibfield  {journal} {\bibinfo  {journal}
  {Proc. R. Soc. London, Ser. A, A282, 318-330}\ } (\bibinfo {year}
  {1964})}\BibitemShut {NoStop}%
\bibitem [{\citenamefont {Knight}\ and\ \citenamefont
  {Rhys}(1959)}]{PdIn_Phasendiagramm}%
  \BibitemOpen
  \bibfield  {author} {\bibinfo {author} {\bibfnamefont {J.}~\bibnamefont
  {Knight}}\ and\ \bibinfo {author} {\bibfnamefont {D.}~\bibnamefont {Rhys}},\
  }\href@noop {} {\bibfield  {journal} {\bibinfo  {journal} {J. Less-Common
  Met. 1 292-303}\ } (\bibinfo {year} {1959})}\BibitemShut {NoStop}%
\bibitem [{\citenamefont {Massalski}(1986)}]{Massalski1986}%
  \BibitemOpen
  \bibinfo {editor} {\bibfnamefont {T.~B.}\ \bibnamefont {Massalski}},\ ed.,\
  \href@noop {} {\emph {\bibinfo {title} {Binary Alloy Phase Diagrams}}},\
  Vol.~\bibinfo {volume} {1}\ (\bibinfo  {publisher} {William W. Scott},\
  \bibinfo {year} {1986})\BibitemShut {NoStop}%
\bibitem [{\citenamefont {Lide}(2005)}]{Handbook}%
  \BibitemOpen
  \bibinfo {editor} {\bibfnamefont {D.~R.}\ \bibnamefont {Lide}},\ ed.,\
  \href@noop {} {\emph {\bibinfo {title} {Handbook of {C}hemistry and
  {P}hysics}}},\ \bibinfo {edition} {85th}\ ed.\ (\bibinfo  {publisher} {{CRC}
  {P}ress},\ \bibinfo {year} {2004-2005})\BibitemShut {NoStop}%
\bibitem [{\citenamefont {Archer}\ and\ \citenamefont
  {Rudtsch}(2003)}]{Archer2003}%
  \BibitemOpen
  \bibfield  {author} {\bibinfo {author} {\bibfnamefont {D.~G.}\ \bibnamefont
  {Archer}}\ and\ \bibinfo {author} {\bibfnamefont {S.}~\bibnamefont
  {Rudtsch}},\ }\href {\doibase 10.1021/je030112g} {\bibfield  {journal}
  {\bibinfo  {journal} {J. Chem. Eng. Data}\ }\textbf {\bibinfo {volume}
  {48}},\ \bibinfo {pages} {1157} (\bibinfo {year} {2003})}\BibitemShut
  {NoStop}%
\bibitem [{\citenamefont {Bhatt}\ and\ \citenamefont
  {Schubert}(1979)}]{Bhatt1979}%
  \BibitemOpen
  \bibfield  {author} {\bibinfo {author} {\bibfnamefont {Y.~C.}\ \bibnamefont
  {Bhatt}}\ and\ \bibinfo {author} {\bibfnamefont {K.}~\bibnamefont
  {Schubert}},\ }\href {\doibase 10.1016/0022-5088(79)90184-X} {\bibfield
  {journal} {\bibinfo  {journal} {J. Less-Common Met.}\ }\textbf {\bibinfo
  {volume} {64}},\ \bibinfo {pages} {P17} (\bibinfo {year} {1979})}\BibitemShut
  {NoStop}%
\bibitem [{\citenamefont {Okawa}\ \emph {et~al.}(2013)\citenamefont {Okawa},
  \citenamefont {Kanou}, \citenamefont {Katagiri}, \citenamefont {Kashiwaya},
  \citenamefont {Kashiwaya},\ and\ \citenamefont {Sasagawa}}]{Okawa2013}%
  \BibitemOpen
  \bibfield  {author} {\bibinfo {author} {\bibfnamefont {K.}~\bibnamefont
  {Okawa}}, \bibinfo {author} {\bibfnamefont {M.}~\bibnamefont {Kanou}},
  \bibinfo {author} {\bibfnamefont {T.}~\bibnamefont {Katagiri}}, \bibinfo
  {author} {\bibfnamefont {H.}~\bibnamefont {Kashiwaya}}, \bibinfo {author}
  {\bibfnamefont {S.}~\bibnamefont {Kashiwaya}}, \ and\ \bibinfo {author}
  {\bibfnamefont {T.}~\bibnamefont {Sasagawa}},\ }\href {\doibase
  10.1016/j.phpro.2013.04.062} {\bibfield  {journal} {\bibinfo  {journal}
  {Physics Procedia}\ }\bibinfo {series} {Proceedings of the 25th
  {International} {Symposium} on {Superconductivity} ({ISS}2012)},\ \textbf
  {\bibinfo {volume} {45}},\ \bibinfo {pages} {101} (\bibinfo {year}
  {2013})}\BibitemShut {NoStop}%
\bibitem [{\citenamefont {Peets}(2014)}]{Peets2014}%
  \BibitemOpen
  \bibfield  {author} {\bibinfo {author} {\bibfnamefont {D.~C.}\ \bibnamefont
  {Peets}},\ }\href {\doibase 10.1088/1742-6596/568/2/022037} {\bibfield
  {journal} {\bibinfo  {journal} {J. Phys. Conf. Ser.}\ }\textbf {\bibinfo
  {volume} {568}},\ \bibinfo {pages} {022037} (\bibinfo {year}
  {2014})}\BibitemShut {NoStop}%
\bibitem [{\citenamefont {Sun}\ \emph {et~al.}(2015)\citenamefont {Sun},
  \citenamefont {Enayat}, \citenamefont {Maldonado}, \citenamefont {Lithgow},
  \citenamefont {Yelland}, \citenamefont {Peets}, \citenamefont {Yaresko},
  \citenamefont {Schnyder},\ and\ \citenamefont {Wahl}}]{Sun2015}%
  \BibitemOpen
  \bibfield  {author} {\bibinfo {author} {\bibfnamefont {Z.}~\bibnamefont
  {Sun}}, \bibinfo {author} {\bibfnamefont {M.}~\bibnamefont {Enayat}},
  \bibinfo {author} {\bibfnamefont {A.}~\bibnamefont {Maldonado}}, \bibinfo
  {author} {\bibfnamefont {C.}~\bibnamefont {Lithgow}}, \bibinfo {author}
  {\bibfnamefont {E.}~\bibnamefont {Yelland}}, \bibinfo {author} {\bibfnamefont
  {D.~C.}\ \bibnamefont {Peets}}, \bibinfo {author} {\bibfnamefont
  {A.}~\bibnamefont {Yaresko}}, \bibinfo {author} {\bibfnamefont {A.~P.}\
  \bibnamefont {Schnyder}}, \ and\ \bibinfo {author} {\bibfnamefont
  {P.}~\bibnamefont {Wahl}},\ }\href {\doibase 10.1038/ncomms7633} {\bibfield
  {journal} {\bibinfo  {journal} {Nat. Commun.}\ }\textbf {\bibinfo {volume}
  {6}},\ \bibinfo {pages} {6633} (\bibinfo {year} {2015})}\BibitemShut
  {NoStop}%
\bibitem [{\citenamefont {Thirupathaiah}\ \emph {et~al.}(2016)\citenamefont
  {Thirupathaiah}, \citenamefont {Ghosh}, \citenamefont {Jha}, \citenamefont
  {Rienks}, \citenamefont {Dolui}, \citenamefont {Ravi~Kishore}, \citenamefont
  {Büchner}, \citenamefont {Das}, \citenamefont {Awana}, \citenamefont
  {Sarma},\ and\ \citenamefont {Fink}}]{Thirupathaiah2016}%
  \BibitemOpen
  \bibfield  {author} {\bibinfo {author} {\bibfnamefont {S.}~\bibnamefont
  {Thirupathaiah}}, \bibinfo {author} {\bibfnamefont {S.}~\bibnamefont
  {Ghosh}}, \bibinfo {author} {\bibfnamefont {R.}~\bibnamefont {Jha}}, \bibinfo
  {author} {\bibfnamefont {E.}~\bibnamefont {Rienks}}, \bibinfo {author}
  {\bibfnamefont {K.}~\bibnamefont {Dolui}}, \bibinfo {author} {\bibfnamefont
  {V.}~\bibnamefont {Ravi~Kishore}}, \bibinfo {author} {\bibfnamefont
  {B.}~\bibnamefont {Büchner}}, \bibinfo {author} {\bibfnamefont
  {T.}~\bibnamefont {Das}}, \bibinfo {author} {\bibfnamefont {V.}~\bibnamefont
  {Awana}}, \bibinfo {author} {\bibfnamefont {D.}~\bibnamefont {Sarma}}, \ and\
  \bibinfo {author} {\bibfnamefont {J.}~\bibnamefont {Fink}},\ }\href {\doibase
  10.1103/PhysRevLett.117.177001} {\bibfield  {journal} {\bibinfo  {journal}
  {Phys. Rev. Lett.}\ }\textbf {\bibinfo {volume} {117}},\ \bibinfo {pages}
  {177001} (\bibinfo {year} {2016})}\BibitemShut {NoStop}%
\bibitem [{\citenamefont {Speil}(1944)}]{DTA_Geschichte}%
  \BibitemOpen
  \bibfield  {author} {\bibinfo {author} {\bibfnamefont {S.}~\bibnamefont
  {Speil}},\ }\href@noop {} {\emph {\bibinfo {title} {\textit{Applications of
  thermal analysis to clays and aluminous minerals}}}},\ \bibinfo {type} {Tech.
  Rep.}\ (\bibinfo  {institution} {Bur. Mines Rep. Invest.},\ \bibinfo {year}
  {1944})\BibitemShut {NoStop}%
\bibitem [{\citenamefont {Chiu}\ and\ \citenamefont
  {Prenner}(2011)}]{Chiu2011}%
  \BibitemOpen
  \bibfield  {author} {\bibinfo {author} {\bibfnamefont {M.~H.}\ \bibnamefont
  {Chiu}}\ and\ \bibinfo {author} {\bibfnamefont {E.~J.}\ \bibnamefont
  {Prenner}},\ }\href {\doibase 10.4103/0975-7406.76463} {\bibfield  {journal}
  {\bibinfo  {journal} {J Pharm Bioallied Sci.}\ }\textbf {\bibinfo {volume}
  {3}},\ \bibinfo {pages} {39} (\bibinfo {year} {2011})}\BibitemShut {NoStop}%
\bibitem [{\citenamefont {Sangster}\ and\ \citenamefont
  {Pelton}(1992)}]{Sangster1992}%
  \BibitemOpen
  \bibfield  {author} {\bibinfo {author} {\bibfnamefont {J.}~\bibnamefont
  {Sangster}}\ and\ \bibinfo {author} {\bibfnamefont {A.~D.}\ \bibnamefont
  {Pelton}},\ }\href {\doibase 10.1007/BF02667557} {\bibfield  {journal}
  {\bibinfo  {journal} {J. Phase Equilib.}\ }\textbf {\bibinfo {volume} {13}},\
  \bibinfo {pages} {291} (\bibinfo {year} {1992})}\BibitemShut {NoStop}%
\bibitem [{\citenamefont {Okamoto}()}]{Okamoto1998_LiN}%
  \BibitemOpen
  \bibfield  {author} {\bibinfo {author} {\bibfnamefont {H.}~\bibnamefont
  {Okamoto}},\ }\href@noop {} {}\bibinfo {note} {{Li-N Phase Diagram, ASM Alloy
  Phase Diagrams Center, P. Villars, editor-in-chief; H. Okamoto and K.
  Cenzual, section editors; http://www1.asminternational.org/AsmEnterprise/APD,
  ASM International, Materials Park, OH, 1998}}\BibitemShut {NoStop}%
\bibitem [{\citenamefont {Elwell}\ \emph {et~al.}(1969)\citenamefont {Elwell},
  \citenamefont {Neate},\ and\ \citenamefont {Smith}}]{Elwell1969}%
  \BibitemOpen
  \bibfield  {author} {\bibinfo {author} {\bibfnamefont {D.}~\bibnamefont
  {Elwell}}, \bibinfo {author} {\bibfnamefont {B.~W.}\ \bibnamefont {Neate}}, \
  and\ \bibinfo {author} {\bibfnamefont {S.~H.}\ \bibnamefont {Smith}},\ }\href
  {\doibase 10.1007/BF01911802} {\bibfield  {journal} {\bibinfo  {journal}
  {Journal of thermal analysis}\ }\textbf {\bibinfo {volume} {1}},\ \bibinfo
  {pages} {319} (\bibinfo {year} {1969})}\BibitemShut {NoStop}%
\bibitem [{\citenamefont {Neate}\ \emph {et~al.}(1971)\citenamefont {Neate},
  \citenamefont {Elwell}, \citenamefont {Smith},\ and\ \citenamefont
  {Agostino}}]{Neate1971}%
  \BibitemOpen
  \bibfield  {author} {\bibinfo {author} {\bibfnamefont {B.~W.}\ \bibnamefont
  {Neate}}, \bibinfo {author} {\bibfnamefont {D.}~\bibnamefont {Elwell}},
  \bibinfo {author} {\bibfnamefont {S.~H.}\ \bibnamefont {Smith}}, \ and\
  \bibinfo {author} {\bibfnamefont {M.~D.}\ \bibnamefont {Agostino}},\ }\href
  {\doibase 10.1088/0022-3735/4/10/017} {\bibfield  {journal} {\bibinfo
  {journal} {J. Phys. E: Sci. Instrum.}\ }\textbf {\bibinfo {volume} {4}},\
  \bibinfo {pages} {775} (\bibinfo {year} {1971})}\BibitemShut {NoStop}%
\bibitem [{\citenamefont
  {{https://scidre.de/product/feedbackfurnace}}()}]{scidrewebsite2}%
  \BibitemOpen
  \bibfield  {author} {\bibinfo {author} {\bibnamefont
  {{https://scidre.de/product/feedbackfurnace}}},\ }\href@noop {}
  {}\BibitemShut {NoStop}%
\end{thebibliography}%

\end{document}